\documentclass[a4paper,fleqn,10pt]{article}
\usepackage{amsmath}
\usepackage{amssymb}
\usepackage{array}
\usepackage{calc}
\usepackage{longtable}
\usepackage{multirow}
\usepackage{textcomp}
\usepackage{gensymb}
\usepackage{rotating}
\usepackage{units}
\usepackage{pstricks}
\usepackage{graphicx}
\usepackage{xspace}
\usepackage{cite}
\usepackage{mcite}
\usepackage{color}
\usepackage[format=hang,labelfont=bf]{caption}
\usepackage{sectsty}
\allsectionsfont{\sffamily}
\subsubsectionfont{\mdseries\itshape\large}
\let\spreprint\empty
\newcommand{\preprint}[1]{\def\spreprint{\protect#1}}
\let\sinstitute\empty
\newcommand{\institute}[1]{\def\sinstitute{\protect#1}}
\makeatletter
\renewcommand{\maketitle}{\begingroup
  \null\thispagestyle{empty}%
    \ifx\spreprint\empty
      \vskip 5ex
    \else
      \flushright\large\spreprint\vskip 2ex
    \fi
    \vskip 5ex
    \flushleft
      {\sffamily\bfseries\huge\@title}\vskip 2ex
      \@author\vskip 2ex
      \ifx\sinstitute\empty
      \else
        {\small\sinstitute}
      \fi
    \vskip 5ex
  \endgroup
}
\makeatother
\renewenvironment{abstract}{\begin{center}
  {\large\sffamily\bfseries Abstract: }
  \begin{minipage}[t]{0.75\textwidth}
}{\end{minipage}\end{center}\vskip 10ex}

\newcommand{\MGfive}{M\protect\scalebox{0.8}{AD}G\protect\scalebox{0.8}{RAPH5\_}aM\protect\scalebox{0.8}{C}@N\protect\scalebox{0.8}{LO}\xspace}
\newcommand{\MCatNLO}{M\protect\scalebox{0.8}{C}@N\protect\scalebox{0.8}{LO}\xspace}
\newcommand{\PowhegBox}{P\protect\scalebox{0.8}{OWHEG}@B\protect\scalebox{0.8}{OX}\xspace}
\newcommand{\OpenLoops}{O\protect\scalebox{0.8}{PEN}L\protect\scalebox{0.8}{OOPS}\xspace}
\newcommand{\GoSam}{G\protect\scalebox{0.8}{O}S\protect\scalebox{0.8}{AM}\xspace}

\newcommand{\Herwig}{H\protect\scalebox{0.8}{ERWIG}\xspace}

\newcommand{\Sherpa}{S\protect\scalebox{0.8}{HERPA}\xspace}

\newcommand{\Rivet}{R\protect\scalebox{0.8}{ivet}\xspace}

\newcommand{\Pythia}{P\protect\scalebox{0.8}{YTHIA}\xspace}

\newcommand{\HepData}{H\protect\scalebox{0.8}{EP}D\protect\scalebox{0.8}{ATA}\xspace}
\long\def\symbolfootnote[#1]#2{\begingroup%
\def\thefootnote{\fnsymbol{footnote}}\footnote[#1]{#2}\endgroup}

\newcommand{\beq}{\begin{equation}}
\newcommand{\eeq}{\end{equation}}
\newcommand{\bal}{\begin{align}}
\newcommand{\eal}{\end{align}}

\preprint{IPPP-14-111\\ DCP-14-222\\ MCNET-14-zzz}
\author{
  Frank Krauss
}
\title{A theory perspective on Top2014}
\institute{
  Institute for Particle Physics Phenomenology,
  Durham University, Durham DH1 3LE, UK
}
\begin{document}
\maketitle
\begin{abstract}
  This is the write-up of the theory keynote talk on the Top2014 conference
  in Cannes, France.
\end{abstract}
\section{Introduction}
It is widely appreciated that the top quark indeed is a very special particle,
for a number of reasons.  First of all, it is the only strongly interacting 
fundamental particle, which does not experience the effect of asymptotic 
freedom: due to its short lifetime it will always decay before the strong
interactions can force it into a bound state.  This in itself makes it a highly
interesting laboratory for precision studies of QCD in the perturbative regime.
Furthermore, and maybe even more importantly, its large mass guarantees the
top quark to play a dominant role in the running of the Higgs boson mass.  This
tight link to the electroweak symmetry breaking sector renders a deeper and
detailed understanding of all of its properties from quantum numbers to 
interaction properties a cornerstone for our understanding of the particle
universe, the fundamental laws of physics at the smallest distances and largest
energies.  In this context it is somewhat amusing to note that its couplings to
the Higgs boson are perturbative although the ratio of the top mass to the 
vacuum expectation value, $m_t/v$ is very close to unity.  This makes a closer 
study of its coupling to the Higgs boson a high priority in top physics, and
in particular the confirmation of the predicted identity of physical top mass 
and its Yukawa coupling $Y_t$ will provide an important test of the Standard 
Model.  This, ultimately provided by precision measurements of the $t\bar{t}H$ 
production rate and distributions, is a very challenging centrepiece of the
the physics programme at the Run II of the LHC.  At the same time, it will
also be important to confirm that the element $V_{tb}$ of the 
Cabibbo--\-Kobayashi--\-Maskawa (CKM) matrix indeed is close to one, as
predicted from the unitarity constraint of the very same matrix.  This can be
achieved by precision studies of single top production at the LHC.  In both
cases, deviations from these relations, $m_t=Y_t$ and $V_{tb}\approx 1$, would
directly signal new physics.  Lastly, due to its large mass, production of top 
quarks in various processes probably is the most notorious background in 
nearly all searches for new physics and thereby a precise understanding of 
processes leading to top quarks in the final state will be hugely important to
find or constrain new physics in direct searches.  

In this contribution I will report on the fairly amazing progress in theory
before the conference, discuss some of the available tools, and will finally 
reflect on the progress on the experimental side.

\section{Theory Progress: High-Precision}
\begin{figure}
  \begin{center}
    \includegraphics[width=0.8\textwidth]{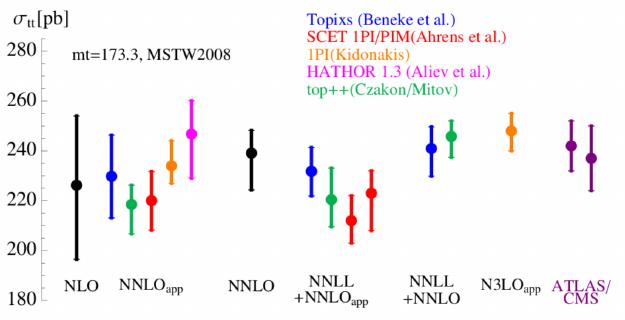}
    \parbox{0.8\textwidth}{\caption{\label{Fig::xsec_mt}
        Various cross sections for top-pair production at the LHC, 
        obtained from different groups in various approximations with the 
        MSTW2008 PDF~\protect\cite{Martin:2009iq}.  Figure taken from a talk by 
        C.~Schwinn at ``top-quark physics day'' at MPI Munich 2014.}}
  \end{center}
\end{figure}

Probably the most notable progress on the theory side is the first complete
calculation of the inclusive top--\-pair cross section to next-to--\-next-to 
leading order (NNLO) accuracy in the strong coupling, reported last year
in~\cite{Czakon:2013goa}.  Their result also forms the basis for a number of 
further calculations, which may further supplement it with the resummation of 
soft logarithms up to next-to--\-next-to leading logarithms or Coulomb effects,
for example~\cite{Beneke:2012wb}.  The results obtained from such calculations 
and some others based on various 
approximations~\cite{Ahrens:2010zv,Aliev:2010zk,Kidonakis:2014isa} 
are compared with each other and data in Fig.~\ref{Fig::xsec_mt}. 
As a consequence of this success it is now possible to extract the top mass
to 3\% precision from the production cross section.  This provides a more than
welcome, simulation independent cross check of the standard determination of 
$m_t$ through top decays or other methods based on kinematics.  

\begin{figure}
  \begin{center}
    \includegraphics[width=0.6\textwidth]{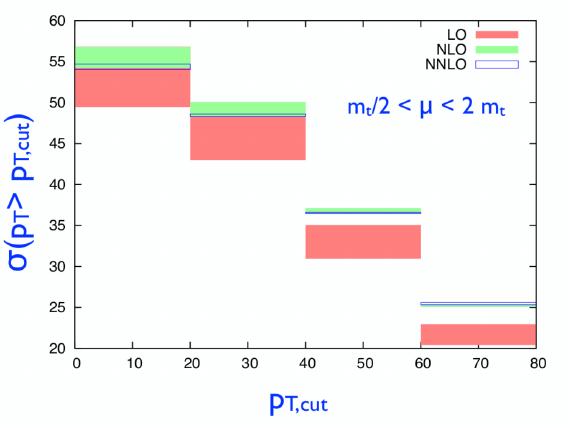}
    \parbox{0.8\textwidth}{\caption{\label{Fig::xsec_singlet}
        Cross section for $t$-channel single-top production at the
        8 TeV LHC, obtained with the MSTW2008 PDF~\protect\cite{Martin:2009iq}, 
        and for different values of the minimal transverse momentum of 
        the top quark, $p_{\rm{T,cut}}$.  Figure taken from a talk by F.~Caola
        at ``top-quark physics day'' at MPI Munich 2014.}}
  \end{center}
\end{figure}

In a similar way, the single-top cross section by now also is known at NNLO 
accuracy.  Focusing on $t$-channel production, which is the dominant channel 
at the LHC, contributing more than 80\% of the overall single-top cross 
section, it is worth noting that the NLO $K$-factor is relatively small, 
roughly of the order of the effect of scale variations on the LO result.  For
a 8 TeV LHC, this correction ranges from 3\% to 18\% in dependence on a cut on 
the top transverse momentum between 0 GeV and 60 GeV.  One may therefore wonder
whether this numerical smallness is an accident.  This has been investigated
in~\cite{Brucherseifer:2014ama}, where only factorisable contributions have 
been taken into account, essentially ignoring cross-talk between the two
coloured lines mediated by real or virtual gluon exchanges.  In this 
approximation the NNLO contribute of the order of about 1-2\% with respect
to the NLO result, ranging from a destructive effect at $p_T>p_{\rm{T,cut}}=0$ 
GeV to a constructive effect at $p_{\rm{T,cut}}=60$ GeV, see also 
Fig.~\ref{Fig::xsec_singlet}.
This result may have been anticipated, as similar processes driven by the 
exchange of colourless particles in the $t$--channel typically usually exhibit 
only relatively small higher-order QCD corrections -- however, to precisely 
quantify such effects is of great value for any precision measurement in this 
channel.  One of the consequences of this result is that by now the CKM element
$V_{tb}$ can be deduced with an accuracy in the order of about 10\%.

In the near future most of these calculations will probably become publicly 
available as parton-level event generators, allowing the production of 
differential distributions, a further great step forward.  Matching with a
subsequent parton shower and therefore enabling event simulation at NNLO
precision is the logical next step, which will most likely happen during 
Run-II.

\section{Progress with Tools}
Turning from the highest precision calculations in top-physics to actual 
state-of-the-art tools ready to be used, it is clear that by now the precision
that can be expected in practically all channels is NLO in the strong 
coupling matched or merged with the parton shower\footnote{
  The first calculations for weak corrections in process involving top quarks
  based on automated tools have been published in~\cite{Frixione:2014qaa}
  for $t\bar{t}H$ final states.
}.  
And while the first such NLO--matched simulation for $t\bar{t}$ production
became available about a decade ago in \MCatNLO~\cite{Frixione:2002ik,
Frixione:2003ei}, by now similar simulations are available for practically all 
processes with up to four particles in the final state, embedding fully 
automated matrix elements at NLO accuracy into parton shower simulations.  At 
the forefront of this development are the \MGfive~\cite{Alwall:2014hca} or  
\PowhegBox~\cite{Nason:2004rx,Frixione:2007vw,Alioli:2010xd} programs, 
interfaced to the \Pythia~\cite{Sjostrand:2007gs,Sjostrand:2014zea} or 
\Herwig~\cite{Bahr:2008pv,Bellm:2013lba} event generators, or more integrated 
solutions such as 
\OpenLoops~\cite{Cascioli:2011va}+\Sherpa~\cite{Gleisberg:2008ta} or
\GoSam~\cite{Cullen:2014yla}+\Sherpa.  

\begin{figure}
  \begin{center}
    \begin{tabular}{cc}
      \includegraphics[width=.4\textwidth]{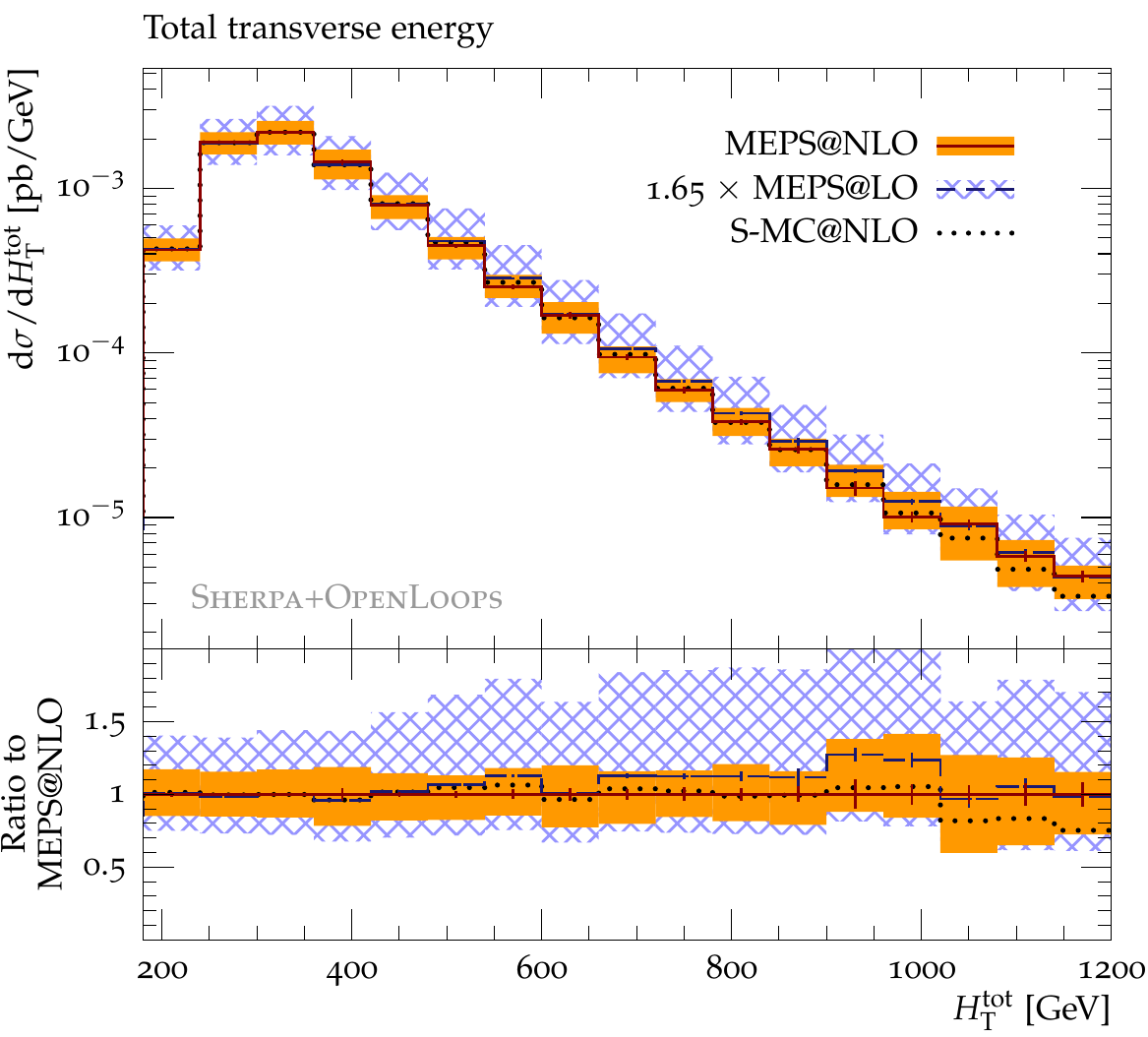} &
      \includegraphics[width=.4\textwidth]{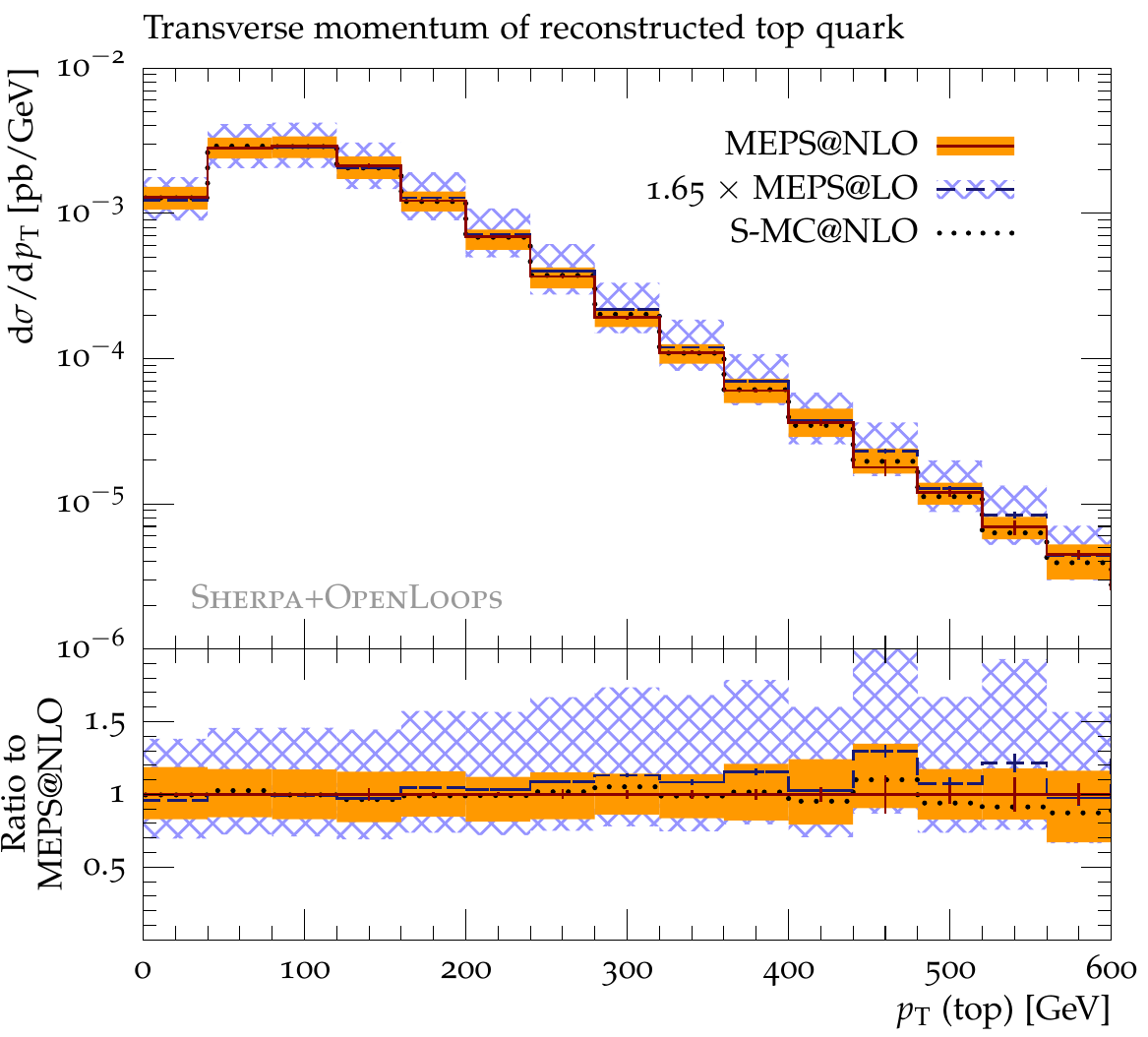}
    \end{tabular}
    \parbox{0.8\textwidth}{\caption{\label{Fig::mc1}
        Differential distributions in $H_T$ (left) and the transverse momentum
        of the top quark (right), obtained with an \OpenLoops+\Sherpa
        simulation~\protect\cite{Hoeche:2014qda}, comparing LO (blue) and 
        NLO--merged (orange) samples and their errors with the S-\MCatNLO 
        simulation in \Sherpa.}}
  \end{center}
\end{figure}

\begin{figure}
  \begin{center}
    \begin{tabular}{cc}
      \begin{minipage}[ht]{.4\textwidth}
      \includegraphics[width=\textwidth]{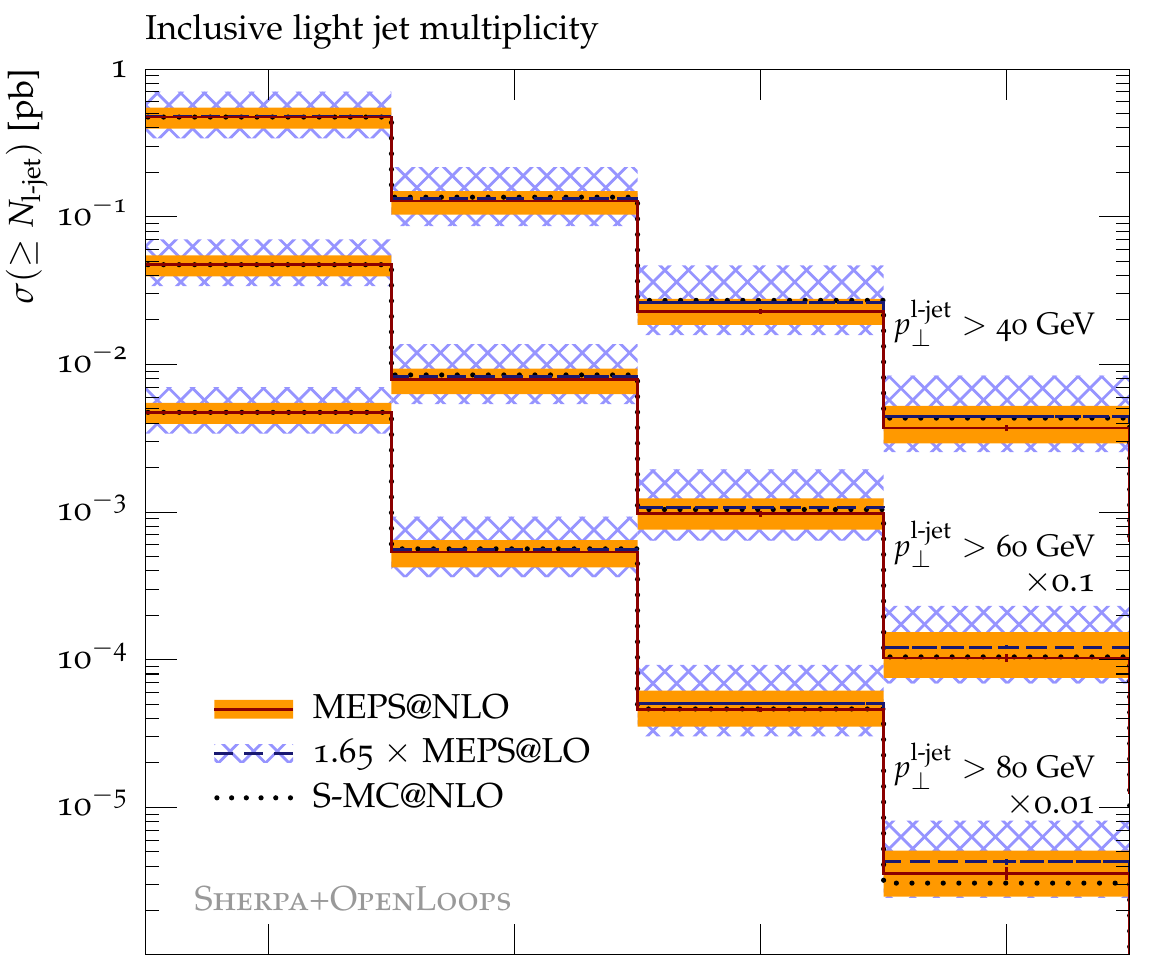}\\
      \includegraphics[width=\textwidth]{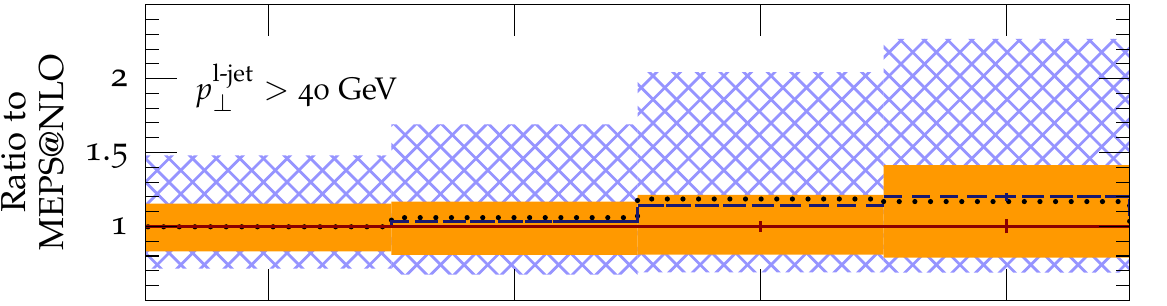}\\
      \includegraphics[width=\textwidth]{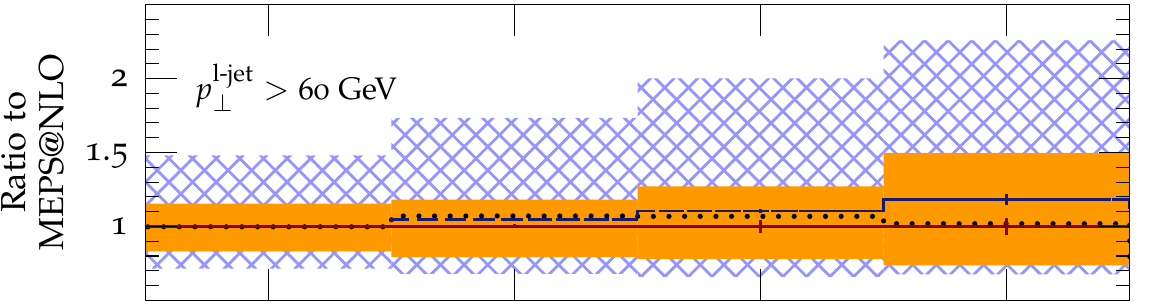}\\
      \includegraphics[width=\textwidth]{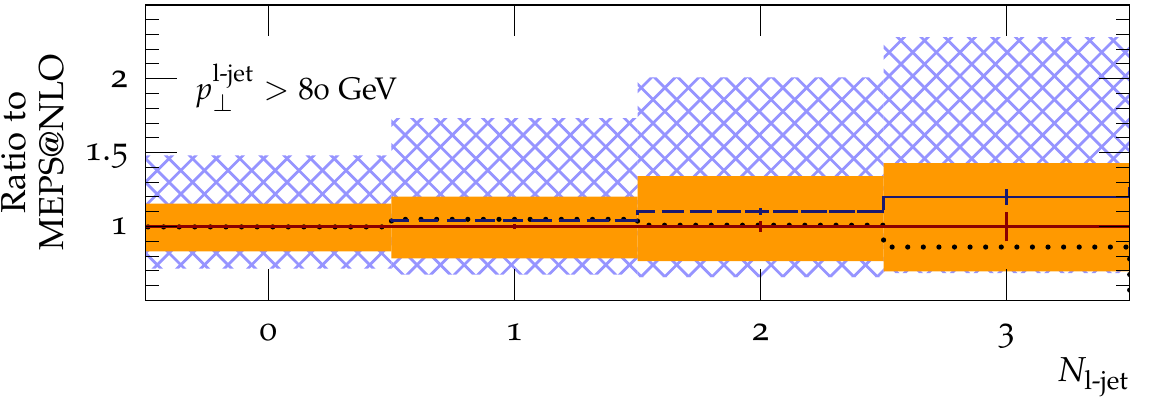}
      \end{minipage} &
      \begin{minipage}[ht]{.4\textwidth}
      \includegraphics[width=\textwidth]{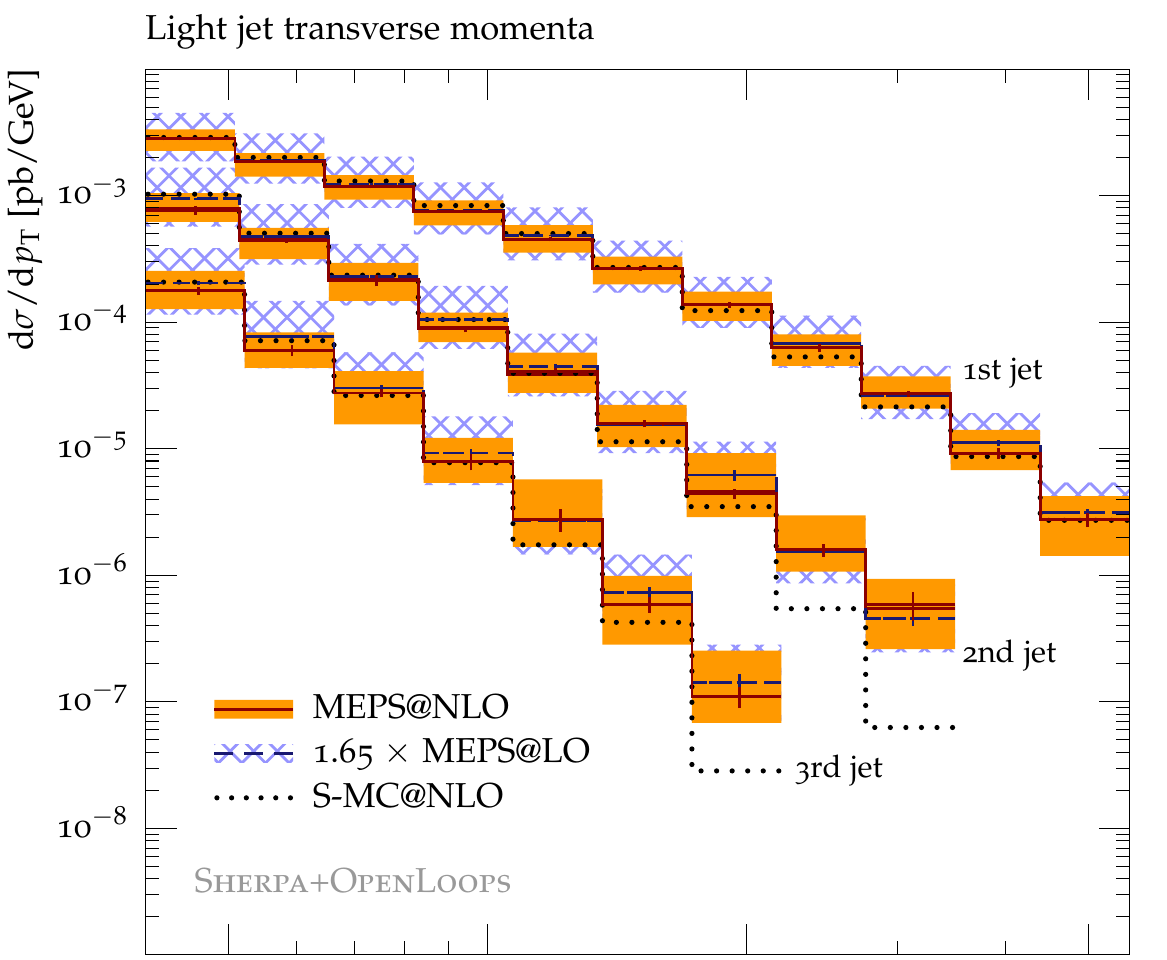}\\
      \includegraphics[width=\textwidth]{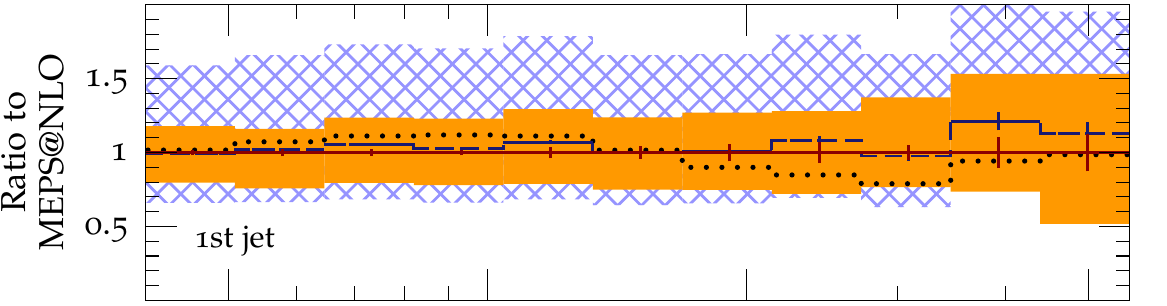}\\
      \includegraphics[width=\textwidth]{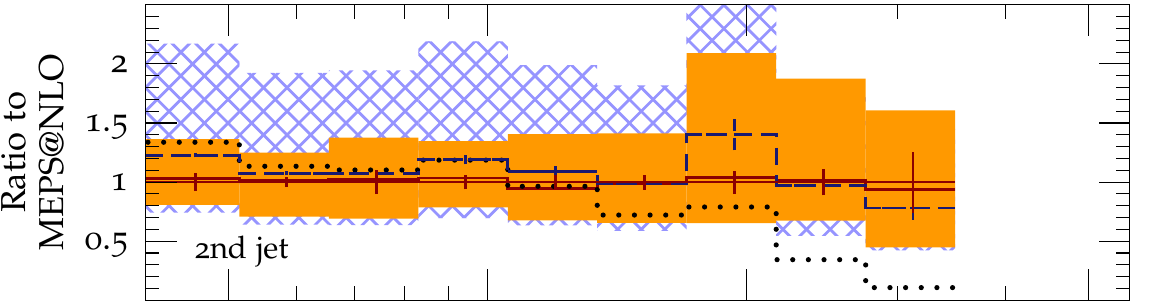}\\
      \includegraphics[width=\textwidth]{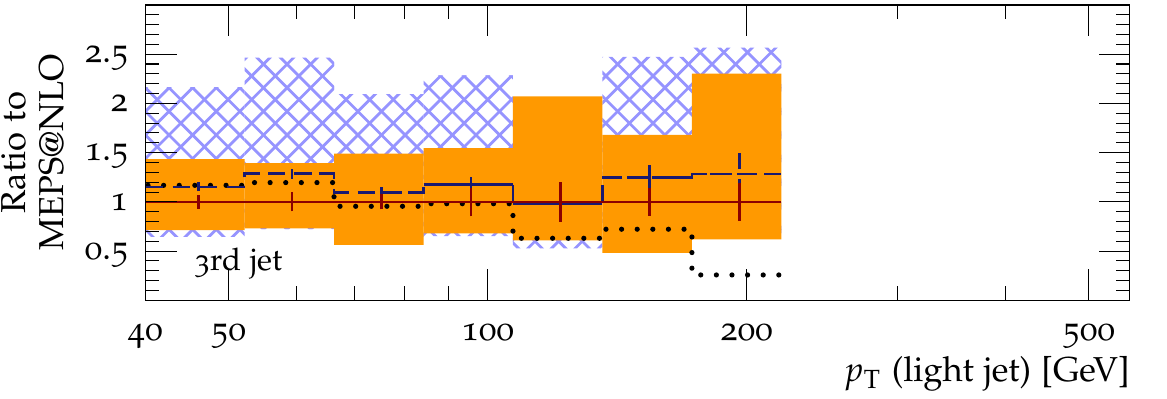}
      \end{minipage} 
    \end{tabular}
    \parbox{0.8\textwidth}{\caption{\label{Fig::mc2}
        Differential distributions in jet multiplicities with various jet 
        transverse momentum thresholds (left) and the transverse momentum
        of the three leading jets, obtained with an \OpenLoops+\Sherpa
        simulation~\protect\cite{Hoeche:2014qda}, comparing LO (blue) and 
        NLO--merged (orange) samples and their errors with the S-\MCatNLO 
        simulation in \Sherpa.}}
  \end{center}
\end{figure}

With these tools it is possible, for instance, to simulate at NLO accuracy 
processes such as top--associated Higgs boson production in $t\bar{t}H$
final states, one of the big challenges for Run-II of the LHC.  This measurement
is necessary to precisely pin down the top-Yukawa coupling.  However, after
applying the usual cuts the backgrounds look very similar to the signal, and
therefore precision simulations will become an indispensable ingredient to every
such measurement.  And while backgrounds such as $t\bar{t}X_S$ production, 
where $X_S$ denotes any singlet such as $Z$, $ZZ$, $W$, etc.\ in principle is
not problematic with the current NLO--matching tools, the production of
additional coloured particles in association with the tops is not so 
straightforward to handle.  This is true for the added intricacies of QCD 
radiation, for example the possible fragmentation of a gluon into a $b\bar{b}$ 
pair, which of course would impact on the number of $b$-tagged jets.  One way 
out is the systematic merging of multijet matrix elements with increasing jet 
multiplicity into one inclusive sample, a procedure that has been pioneered 
in~\cite{Catani:2001cc,Lonnblad:2001iq,Mangano:2001xp,Krauss:2002up}
about a decade ago for such towers of LO processes.  Recently, the same
logic has also been extended to multijet merging of NLO matrix elements in
various schemes~\cite{Gehrmann:2012yg,Hoeche:2012yf,Frederix:2012ps,
  Lonnblad:2012ix,Platzer:2012bs}.
Examples for results obtained with this procedure, for the case of 
$t\bar{t}+$jets production, taken from~\cite{Hoeche:2014qda}, are displayed
in Figs.~\ref{Fig::mc1} and \ref{Fig::mc2}.  While inclusive observables such
as the ones displayed in the former do not show any improvement of the 
NLO--merged sample over a regular NLO--matched one, the more exclusive 
observables exhibited in the latter do.
In the next months such simulation tools will need to be further scrutinised
and compared.  Systematic uncertainties such as the one related to the choice
of the merging scale or the renormalisation and factorisation scale must be
further quantified.  In addition, it is time to define meaningful ways to
systematically evaluate the uncertainties stemming from the parton shower and 
the non--\-perturbative parts of the simulation.

Finally, concerning tools it is interesting to note that with Run-II the 
boosted regime of top production will become increasingly important: ultimately
the LHC will become a factory of highly boosted objects, see for example
Tab.~\ref{Tab::boosted} and transverse momenta at the TeV scale will be 
copiously produced and frequently studied.  There is a large and apparently 
ever increasing number of tools and procedures to extract information out of 
these topologies, but it is probably important now to understand their 
similarities and differences in more detail, to fully appreciate their 
relative virtues, and to work out their respective systematic uncertainties.
\begin{table}[h]
  \begin{center}
    \begin{tabular}{|l||c|c|c|}
      \hline
      $N_{t\bar{t}}$      & Tevatron Run-II & LHC 2012 & LHC design\\
      in three regimes & 10 fb$^{-1}$, 1.96 TeV & 20 fb$^{-1}$, 8 TeV & 
      300 fb$^{-1}$, 13 TeV \\
      \hline
      inclusive          & 57k & 2.6M & 155M\\
      $M_{t\bar{t}}>1$ TeV  & 25  & 30k &  3M\\
      $M_{t\bar{t}}>2$ TeV  &  0  & 300 &  47k\\
      \hline
    \end{tabular}
    \parbox{0.8\textwidth}{\caption{\label{Tab::boosted}
        Number of expected $t\bar{t}$ events in three different regimes 
        at three different collider setups.}}
  \end{center}
\end{table}

\section{Experimental Progress}
Naturally, it is very hard and possibly not fully adequate to also comment on
the progress on the experimental side -- I will try to formulate a few 
thought and concerns nevertheless.  

After its discovery there has been a large number of ground-breaking 
measurements of top quark properties, some of which have large impact on our 
understanding of the inner dynamics and consistency of the Standard Model.
The most important example obviously is the precise determination of the
top mass, which by now has errors at the GeV scale only.  The only standing
concern in most of the measurements to date is that they are quite often
heavily based on MC event generators and, for example, templates obtained 
from them.  This of course immediately poses the question in how far the
measured value is a very MC specific mass, dependent on details of how parton
showering etc.\ is being implemented.  Fortunately, both the theory and the 
experimental community have picked up on this issue, and consequently there
is a lot of progress on both sides, by fixing the MC scheme the mass is defined
in -- very close to the on-shell scheme -- and by moving to more and better
observables, such as the total cross section or similar, which allow a more
MC--\-independent determination of the top mass.  As there have been a number 
of talks on this subject during the conference, for example the ones by 
Hoang~\cite{Hoang:2014oea} and Adomeit~\cite{Adomeit:2014yna}, I will not 
further elaborate here.  

Turning to inclusive pair-production cross sections, the sheer amount of
results collected by the Tevatron and LHC experiments in the past years is
nothing but staggering.  It is fairly satisfying to see the by far and large 
very good consistency of the measured cross sections among each other and
also their typically excellent agreement with theory, see for example the talks
presented by Shabalina and Brochero Cifuentes.  It is fair to say that this 
certainly is a job well done and continuing to be well done!

Turning to more differential distributions, the picture is not quite as 
convincing.  Up to this conference, and with a few notable exceptions such
as for instance the measurement of rapidity gaps~\cite{ATLAS:2012al}, or of 
jet multiplicities~\cite{Chatrchyan:2014gma} in top-pair production,
quite a lot of the results are to be taken with a grain of salt: Quite
often they are either over--\-corrected, for instance by extrapolating to
full acceptance detectors or by correcting to top-quark partons, or 
under--\-corrected by not accounting for detector effects.  

Take as an example the inclusive production of top-pairs.  It appears as if 
there is a simple pattern in how the currently used tools perform with respect
to data: While by far and large they appear to agree very well with the 
invariant mass spectrum of the pair, $m_{t\bar{t}}$, the transverse momentum
distribution of the individual tops, $p_\perp^{(t)}$, seems to be a bit tricky.
It would be great if the theory community could check these effects and try 
to find possible issues related to the simulation.  This, however, is not quite
possible, as quite often the data are extrapolated to a full acceptance 
(``$4\pi$'')detector and to top partons.  By correcting in this way, the data
already are modified by a simulation tool and it is very hard or probably even 
impossible to exactly figure out how this modified the results.  It is 
therefore crucial that the experiments {\em report} their data, corrected for 
detector effects, on the particle level, i.e.\ on the basis of physical 
objects, within the actual acceptance of the detector.  Of course, a further
{\em interpretation} of the results, by comparing on the parton level with 
higher-order calculations is more than welcome then.  

\begin{figure}[h]
  \begin{center}
    \includegraphics[width=0.6\textwidth]{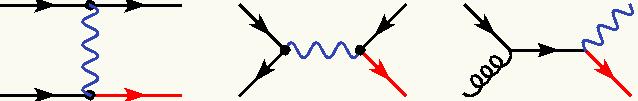}    
    \parbox{0.8\textwidth}{\caption{\label{Fig::single_top_Feyn}  
        Feynman diagrams for single-top production at LO.}}
  \end{center}
\end{figure}

\begin{figure}[h]
  \begin{center}
    \includegraphics[width=0.6\textwidth]{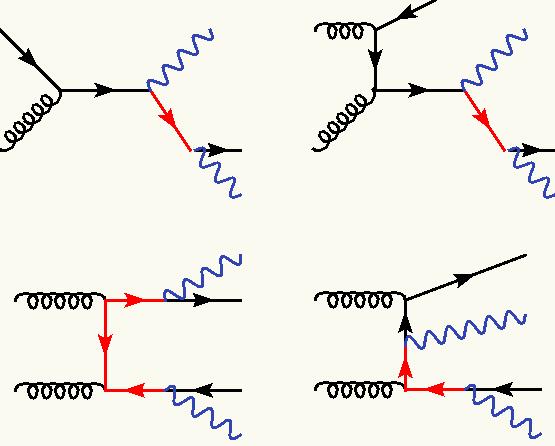}
    \parbox{0.8\textwidth}{\caption{\label{Fig::Wt_Feyn}  
        Feynman diagrams for $Wt$ production, including the decay of the
        top quark: LO contribution (top left) and NLO contributions (top
        right and bottom).}}
  \end{center}
\end{figure}

This is even more important for channels which are not very clearly defined.
As an example for this consider single-top production.  Common lore has it
that this process proceeds along three channels: $t$-channel, $s$-channel, 
and $Wt$ production, see Fig.\ \ref{Fig::single_top_Feyn}.  
While this works nicely at leading order, this discrimination breaks down 
at higher orders; adding the decay of the top quark, at NLO accuracy 
$Wt$--associated production will have contributions e.g.\ from
top pair production, as shown in Fig.\ \ref{Fig::Wt_Feyn}.
This shows that such a differentiation makes sense at LO level only, where 
$Wt$ production contributes about 20\% of the total single-top cross section.  
To me, trying to ``measure'' $Wt$ production cross sections sounds a bit like 
a waste of time there; instead it seems more reasonable to define physical 
objects in the full $b\bar{b}W^+W^-$ final states, for instance one could
identify regions of double-, single- or non-resonant $bW$ production, and 
(rightly) interpret the resonances as top quarks.  Reporting measurements 
based on fiducial cross sections would then ensure that different 
interpretations, like for instance top production in different channels, become
feasible.  

A simple lesson from this example is that in presenting the data -- and 
especially those for differential cross sections -- it must be guaranteed
that they can be interpreted to all order of perturbation theory; this is the
only way to ensure that the gruelling efforts in and the amazing successes of 
the higher-order calculations discussed in the first part of the talk are not 
in vain.  Of course, this does not forbid that such data, based on physical
object definitions and the fiducial volume of the detector are extrapolated to 
the parton level at full coverage to be directly compared to such calculations!
This comparison certainly is an integral part of the interpretation of the 
data.

I am more than happy that during the conference a number of talks presented
results where exactly the procedure outlined above -- data based on physical 
objects in the fiducial volume -- has been followed.  One example for this
is the pseudo-top method reported for instance in the talk by Katzy.

\section{Concluding remarks}
Instead of a summary I would like to take the opportunity to briefly discuss
certain aspects of data preservation.  This is based on the very real 
possibility that the LHC is {\bf the experiment} of our lifetimes.  This 
implies that, in a lab-based and controlled environment, we explore physics at 
the largest energies ever accessible to us\footnote{
  The slightly sad corollary is that going to Run II is the last energy upgrade
  we may experience.}.
Clearly, if this is the case we should do our utmost to make this experiment
count.  This means that we owe it to us and to future generations to make
our results as reproducible as possible.  But how can this be achieved?
First of all, it is obvious that nobody but the current experimenters is better
placed to understand the detector, which implies that any data not unfolded
during the lifetime of an experiment will never be unfolded.  In many cases
this will result in the data effectively being unusable, essentially lost.
Therefore: please, unfold your measurements to the particle level.  Furthermore,
my feeling is that it is probably simpler to present, understand and possibly
migrate the detailed knowledge of the experimental environment of an analysis,
the cuts employed etc., when it is based on code rather than text.  Text nearly 
always tends to have a slightly approximate character, which is not the case 
with computer code.  A tool like \Rivet~\cite{Buckley:2010ar} comes to my mind 
here, but any other publicly available and documented software is as good.  It 
just must be linked to the data, stored on a database such as \HepData.  
Similar reasoning also holds true for the Monte Carlo tools -- in order to 
allow reproducibility it would be great if the exact version of the used
code(s), the run card(s) or similar would be tagged and linked to the data.
This, together with only using publicly available code would allow future
generations to reproduce our results.  In addition it would offer them
a great way to learn from our errors and mistakes.  

\section*{Acknowledgements}
First of all I would like to thank the organisers for a wonderful conference 
with a great atmosphere -- it was a real pleasure being in Cannes!  I also
gratefully acknowledge financial support by the EU Initial Training Networks
``MCnet''  (PITN-GA-2012-315877) and ``HiggsTools'' (PITN-GA-2012-316704) and
by the ERC Advanced Grant MC@NNLO (340983).


\end{document}